\documentclass[jkps,preprint,fleqn,showpacs,showkeys]{revtex4}
\usepackage{graphicx}
\usepackage{amssymb}
\usepackage{amsmath}
\usepackage{bm}
\begin{document}
\setcounter{page}{0}
\title[]{Dirac Coupled Channel Analyses of the 2$^-$ Gamma Vibrational
band excitation in $^{20}$Ne}
\author{Sugie \surname{Shim}}
\email{shim@kongju.ac.kr}
\thanks{Fax: +82-41-850-8489}
\affiliation{Department of Physics, Kongju National University, Kongju 314-701}

\date[]{Received 2014}

\begin{abstract}
Dirac coupled channel analyses are performed using
optical potential model for the high-lying excited states that belong to the 2$^-$ gamma vibrational
band at the 800 MeV unpolarized proton inelastic scatterings from $^{20}$Ne. The first order vibrational
collective models are used to obtain the transition optical potentials to
describe the high-lying excited vibrational collective states and Lorentz-covariant
scalar and time-like vector potentials are used as direct optical potentials. The complicated
Dirac coupled channel equations are solved phenomenologically to reproduce the
differential cross sections data by varying the optical potential and deformation parameters using minimum chi-square
method. It is found that relativistic Dirac coupled channel calculation could describe the excited states of the 2$^-$ gamma vibrational band in $^{20}$Ne much better than the nonrelativistic coupled channel calculation, especially for the 2$^-$ and 3$^-$ states of the band.
It is shown that the multistep excitation process via channel coupling with the $3^-$ state is essential to describe the $2^-$ state excitation and pure direct transition from the ground state is dominant for the 3$^-$ state excitation of the 2$^-$ gamma vibrational band in $^{20}$Ne.

\end{abstract}



\maketitle

\section{INTRODUCTION}

Relativistic treatment of nuclear reactions based on use of the Dirac equation have proved to be very successful, in particular for the description of elastic and inelastic nucleon-nucleus scatterings\cite{1,2,3,4,5,6}.
It has been shown that considerable improvements are obtained in the coupled channel calculations for the intermediate energy proton inelastic scatterings from spherically symmetric nuclei and a few deformed nuclei using Dirac phenomenology compared to the conventional nonrelativistic calculations based on Schr\"{o}dinger equation\cite{4,5,6,7,8,9,10,11}.
In this work we performed Dirac coupled channel analyses for the high-lying excited states of an s-d shell nucleus $^{20}$Ne that belong to the 2$^-$ gamma vibrational
band($K^\pi = 2^- $) at the intermediate energy proton inelastic scatterings. We use optical potential model, employing S-V model where only scalar and time-like vector potentials are considered. Woods-Saxon shape is used for the geometry of the direct optical potentials.
In order to accommodate the collective  motion of the excited deformed nucleus considering the high-lying excited states that belong to 2$^-$ gamma vibrational
band in $^{20}$Ne, the first order vibrational collective model is used to obtain the transition optical potentials. Possible 2$^-$, 3$^-$ and 5$^-$ excited states of the $K^\pi = 2^- $ octupole band are considered in the calculation.
The complicated Dirac coupled channel equations are solved using a computer program called ECIS\cite{12} where Dirac optical potential and deformation parameters are determined using sequential iteration method. The channel-coupling effect of multistep process for the excited states of the 2$^-$ gamma vibrational
band is investigated.  The calculated results are analyzed and compared with the experimental data and those of nonrelativistic approaches.

\section{Theory and Results}

Dirac coupled channel calculations are performed phenomenologically for the high-lying excited states that belong to the 2$^-$ gamma vibrational
band at the unpolarized proton inelastic scatterings from $^{20}$Ne using optical potential model and the first order vibrational collective model.
Since $^{20}$Ne is one of the spin-0 nuclei, only scalar, time-like vector and tensor
optical potentials can survive\cite{13,14}, hence the relevant Dirac equation for the elastic scattering from the nucleus is given as
\begin{equation}
[\alpha \cdot p + \beta ( m + U_S ) - ( E - U_0- V_c )
 + i \alpha \cdot  \hat{r} \beta U_T ] \Psi(r) = 0
\label{e1}
\end{equation}
Here, $U_S$ is a scalar potential, $U_0 $ is a time-like vector potential,  $U_T$ is a tensor potential, and $V_c $ is the Coulomb potential.
However, it is true that pseudo-scalar and
axial-vector potentials may also be present in the equation when we consider inelastic
scattering, depending on the  model assumed. In this work, it is assumed that appropriate transition potentials can be obtained by deforming the direct potentials that describe the elastic
channel reasonably well\cite{15}.
Transition potentials are obtained by assuming that they are proportional to the first-order derivatives of the diagonal potentials. The scalar and the time-like vector potentials are used as direct potentials in the calculation. Even though tensor potentials are always present due to the interaction of the anomalous magnetic moment of the projectile with the charge distribution of the target, they have been found to be always very small compared to scalar or vector potentials\cite{4}. Hence, they are neglected in this calculation.
In the vibrational model of ECIS, the deformation of the nuclear surface is written using the Legendre polynomial expansion method,
\begin{equation}
R(\theta, \phi ) = R_0 ( 1+ \sum_{\lambda \mu } \beta^i _\lambda Y^* _{\lambda \mu } (\theta, \phi ) )
\label{e2}
\end{equation}
with $R_0$ the radius at equilibrium, $\beta$ is a deformation parameter and $\lambda$ is the multipolarity.
  The transition potentials are given by
\begin{equation}
U_i ^\lambda = \sum_\mu \frac{\beta^i _\lambda R_i }{(2\lambda +1)^{1/2}} \frac{dU_i (r)}{dr } Y^* _{\lambda \mu} (\Omega)
\label{e3}
\end{equation}
where the subscript $i$  refer to the real and imaginary scalar or vector potential and $R$ is the radius parameter of Woods-Saxon shape. The real and imaginary deformation parameter $\beta_\lambda$'s are taken to be equal for the given  potential type so that two deformation parameters of $\beta_S$ and $\beta_V$ are determined for each excited state.

In order to compare the calculated results with those of the previous nonrelativistic calculations, Dirac equation is reduced to the Schr\"{o}dinger-like second order differential equation by considering the upper component of Dirac wave function and the effective central and spin-orbit optical potentials are obtained\cite{4}.
It should be noted that one of the merit of relativistic approach based on Dirac equation instead of using the nonrelativistic approach based on Schr\"{o}dinger equation is that the spin-orbit potential appears naturally in Dirac approach when Dirac equation is reduced to the Schr\"{o}dinger-like second order differential equation, while the spin-orbit potential should be put by hand in the nonrelativistic Schr\"{o}dinger approach.

The experimental data of the differential cross sections are obtained from Ref. 18 for the 800 MeV unpolarized proton inelastic scatterings from $^{20}$Ne. The high-lying excited states of 2$^-$ gamma vibrational band, the $2^-$(4.97MeV), $3^-$(5.62MeV) and $5^-$(8.45MeV) states are considered and assumed to be collective vibrational states in the calculation.

First, the 12 parameters of the diagonal scalar and vector potentials in Woods-Saxon shapes are determined phenomenologically by fitting the elastic scattering differential cross section experimental data.
The calculated results of the 12-parameter search are shown as short-dash lines in Fig. 1 and it is found that the differential cross section experimental data are reproduced quite well. The calculated optical potential parameters of Woods-Saxon shape for the 800 MeV proton elastic scatterings from $^{20}$Ne are shown in Table 1, showing almost the same values with the case where polarized proton scatterings from $^{20}$Ne are considered\cite{16}.
It is noted that the real scalar potentials and the imaginary vector potentials are turned out to be large and negative, and the imaginary scalar potentials and the real vector potentials are turned out to be large and positive, showing the same pattern as in the spherically symmetric nuclei\cite{4}.
\begin{figure}
\includegraphics[width=10.0cm]{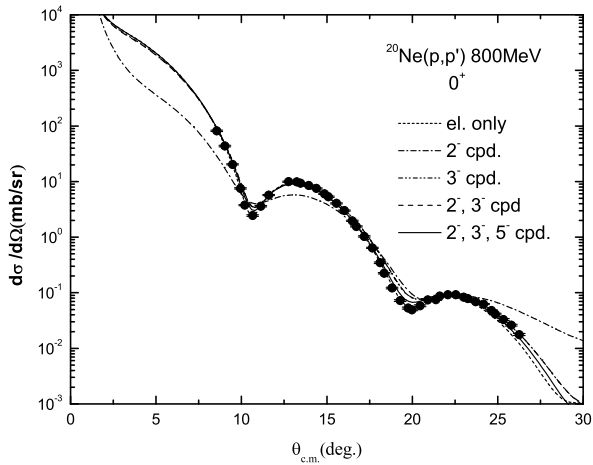}
\caption[0]{Differential cross section of the ground state for 800 MeV p +  $^{20}$Ne scattering. Short dash line, dash-dot, dash-dot-dot, dashed and solid lines represent the results of Dirac phenomenological calculation where only the ground state is considered, where the ground and $2^-$ states are coupled, where the ground and $3^-$ states are coupled, where the ground, $2^-$  and $3^-$ states are coupled, and where all four states of 2$^-$ gamma vibrational band are coupled, respectively.}
\label{fig1}
\end{figure}

\begin{figure}
\includegraphics[width=10.0cm]{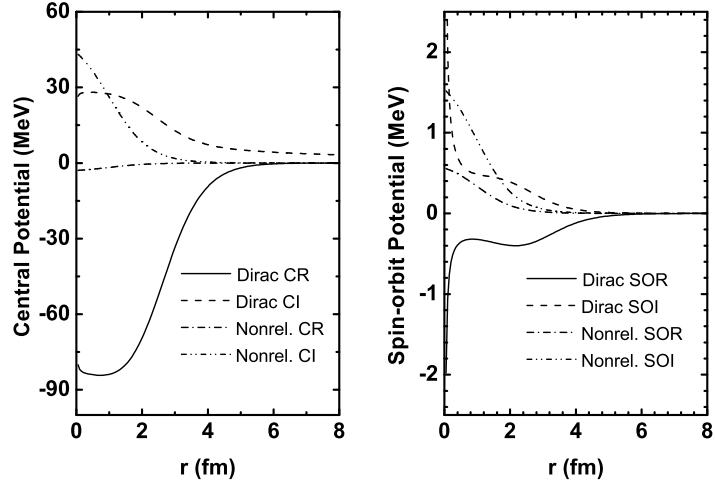}
\caption[0]{Comparison of Dirac effective central and spin-orbit potentials of $^{20}$Ne with those of nonrelativistic calculation. CR, CI, SOR and SOI mean central real and imaginary, spin-orbit real and imaginary potentials, respectively.}
\label{fig2}
\end{figure}

\begin{table}
\caption{The calculated phenomenological optical potential parameters of Woods-Saxon shape for the 800 MeV unpolarized proton elastic scatterings from $^{20}$Ne.}
\begin{ruledtabular}
\begin{tabular}{cccccccc}
   Potential      &   strength(MeV)   & radius(fm)  & diffusiveness(fm)  ~ \\
   \hline
 Scalar  & -200.7     & 2.677 & 0.7578        ~ \\
 real    &      &     &            ~ \\ \hline
 Scalar  & 127.2     & 1.859 & 0.7802        ~ \\
 imaginary    &      &     &            ~ \\ \hline
 Vector  & 113.0    & 2.693 & 0.7419         ~ \\
 real    &      &     &           ~ \\ \hline
 Vector  & -107.4     & 2.409 & 0.6781       ~ \\
 imaginary  &      &     &             ~ \\
\end{tabular}
\end{ruledtabular}
\label{table1}
\end{table}

In Fig. 2, Dirac effective central and spin-orbit potentials of $^{20}$Ne are compared with those of nonrelativistic calculations\cite{17} and it is found that surface-peaked phenomena are observed for the real parts of the effective central potentials(CR), as in the case of $^{24}$Mg\cite{11}.  The strength of the real effective central potential of the Dirac approach turned out to be large, giving about -80MeV at the center of the nucleus, compared to that of nonrelativistic Schr\"{o}dinger approach which was about -3.5MeV.
The surface-peaked phenomena are also observed for the real parts(SOR) of spin-orbit potentials, indicating spin-orbit interaction is surface-peaked interaction. However, the peak position of the real effective spin-orbit potential is observed to be slightly different from that of the real effective central potential, located a little more outside of the nucleus.
Surface-peaked shape cannot be obtained in the nonrelativistic calculations because they use Woods-Saxon shape for both central and spin-orbit potentials.

Next, by including one excited state, the $2^-$ state or the $3^-$ state of the 2$^-$ gamma vibrational band in addition to the ground state, six-parameter searches are performed starting from the obtained 12 parameters of direct optical potentials. Here, the six parameter means the four potential strengths; they are the scalar real and imaginary potential strengths and the vector real and imaginary potential strengths, keeping the potential geometry unchanged, and two deformation parameters, $\beta_S$ and $\beta_V$, of the included excited state. Here, the four optical potential strengths obtained by fitting to the elastic scattering data are varied since the channel coupling of the excited states to the ground state should be included in the inelastic scattering calculation.
Dirac coupled channel equations are solved to obtain the best fitting parameters to the experimental data using minimum $\chi^2$ method numerically.
As a next step, eight-parameter searches are performed by including the $2^-$ and $3^-$ excited states in addition to the ground state.
Finally, ten-parameter searches are performed by considering all four states, the ground, $2^-$, $3^-$ and $5^-$ states all together in the calculation and the results are compared with those of the calculation where only the ground and one or two excited states are coupled, in order to investigate the effect of the channel coupling between the excited states.

\begin{figure}
\includegraphics[width=10.0cm]{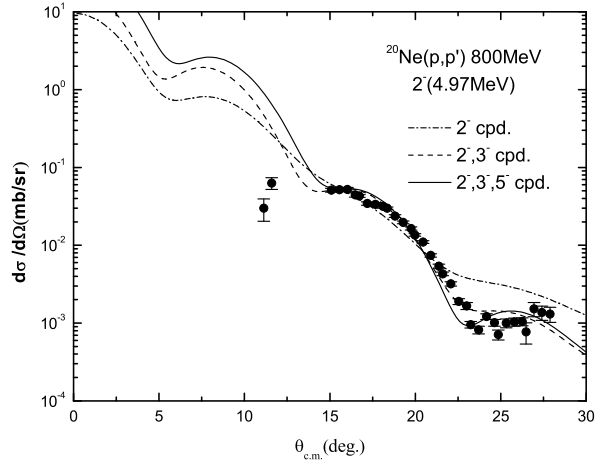}
\caption[0]{Differential  cross section of the $2^- $ state for 800 MeV p +  $^{20}$Ne scattering. Dash-dot line, dashed, and solid lines represent the results of Dirac coupled channel calculation where the ground and $2^-$ states are coupled, where the ground, $2^-$ and $3^-$ states are coupled, and where all three states of 2$^-$ gamma vibrational band are coupled, respectively.}
\label{fig3}
\end{figure}

\begin{figure}
\includegraphics[width=10.0cm]{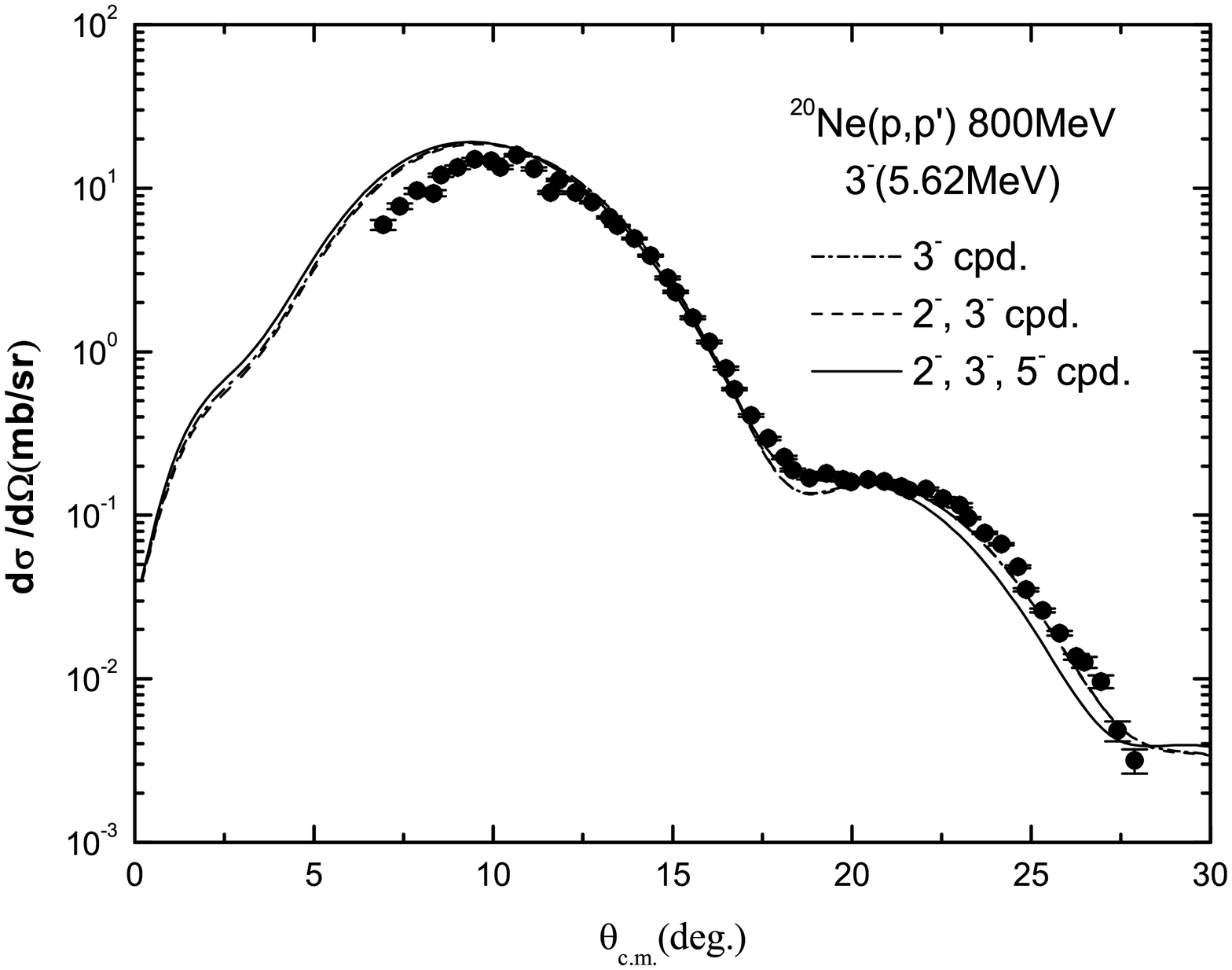}
\caption[0]{Differential  cross section of the $3^- $ state for 800 MeV p +  $^{20}$Ne scattering. Dash-dot line, dashed, and solid lines represent the results of Dirac coupled channel calculation where the ground and $3^-$ states are coupled, where the ground, $2^-$ and $3^-$ states are coupled, and where all three states of 2$^-$ gamma vibrational band are coupled, respectively.}
\label{fig4}
\end{figure}

\begin{figure}
\includegraphics[width=10.0cm]{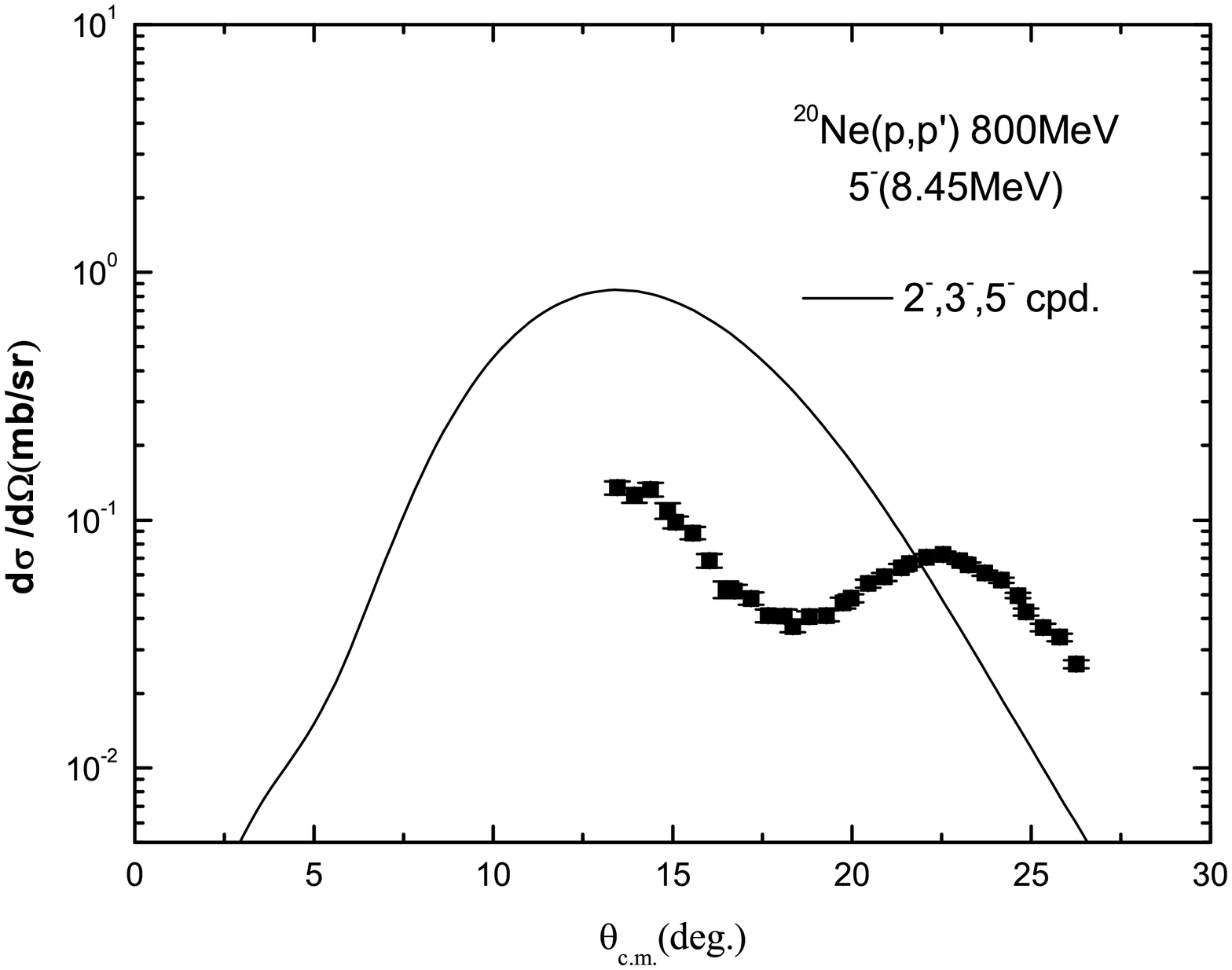}
\caption[0]{Differential  cross section of the $5^- $ state for 800 MeV p +  $^{20}$Ne scattering. Solid lines represent the results of Dirac coupled channel calculation where all three states of 2$^-$ gamma vibrational band are coupled.}
\label{fig5}
\end{figure}

The results of the coupled channel calculations for the ground state are given in Fig. 1. It is shown that most of them reproduce the elastic experimental data pretty well except the case where only $2^-$ state is coupled. For only the $2^-$ state coupled case, even the ground state is not well described. It seems due to that the $2^-$ state is excited only by transitions which involve at least two steps, and not from the ground state directly\cite{18}.
The calculated observables of the $2^-$ state are shown in Fig. 3. The dash-dot, dashed and the solid lines represent the results of the calculation where the ground and $2^-$ states are coupled, where the ground, $2^-$ and $3^-$ states are coupled, and where all four states are coupled, respectively. It is clearly shown that the agreement with the $2^-$ differential cross section data is improved significantly by adding the $3^-$ state.  It can be interpreted as that it is essential to include the two-step transition process via the channel coupling with the $3^-$ state in order to describe the $2^-$ state excitation\cite{9, 18}. When the coupling with the $5^-$ state is added in the calculation, it is observed that the third minimum of the diffraction pattern went downward a little.
Figure 4 shows the calculated results for the $3^-$ state excitation. The agreement with the $3^-$ state experimental data turned out to be very good for the both cases of where the $2^-$ and $3^-$ states are included and all four states are included, showing much better agreement with the experimental data comparing to the results of the nonrelativistic calculation\cite{18}. Even for the case where only the $3^-$ state is coupled, the agreement with the experimental data turned out to be pretty good, indicating the excitation by direct transition from ground state is dominant for this state.

\begin{table}
\caption{ Comparison of the deformation  parameters for the excited states of the 2$^-$ gamma vibrational band in $^{20}$Ne at the 800 MeV proton inelastic scatterings with those of nonrelativistic calculations\cite{18}. Potential strengths are ordered as scalar real and imaginary, vector real and imaginary, downward from the top.
$\chi_{e}^{2}$/N and $\chi_{i}^{2}$/N are reduced chi-squares for elastic and inelastic
cross section fitting results, respectively.}
\begin{ruledtabular}
\begin{tabular}{ccccccccccc}
         &   E   & $\chi_{e}^{2}$/N & $\chi_{i}^{2}$/N & $\beta_{S}$ & $\beta_{V}$ & $\beta_{NR}$ & Potential ~ \\
         & (MeV) &      &         &            &            &             & (MeV) ~ \\ \hline
 Elastic &      & 16.8 &     &      &      &      & -200.7  ~ \\
 only    &      &     &     &      &      &      &  127.2  ~ \\
         &      &     &     &      &      &      &   113.0  ~ \\
         &      &     &     &      &      &      & -107.4 ~ \\ \hline
 2$^-$   & 4.97 & 273 & 56.0 & -.050 & -.044 &      & -271.7  ~ \\
  cpd.   &      &     &     &      &      &      &   526.7  ~ \\
         &      &     &     &      &      &      &   168.5  ~ \\
         &      &     &     &      &      &      &  -213.6  ~ \\ \hline
 3$^-$   & 5.62 & 53.8 & 80.5 & .445 & .452 &      & -189.1  ~ \\
  cpd.   &      &     &     &      &      &      &   168.3  ~ \\
         &      &     &     &      &      &      &   113.6  ~ \\
         &      &     &     &      &      &      &  -118.1  ~ \\ \hline
 2$^-$   & 4.97 & 58.2 & 18.1 & .074 & .076 &      & -200.2  ~ \\
 3$^-$   & 5.62 &     & 75.4 & .446 & .450 &      &   173.3  ~ \\
  cpd.   &      &     &     &      &      &      &   119.8  ~ \\
         &      &     &     &      &      &      &  -118.3  ~ \\ \hline
 2$^-$,  & 4.97 & 20.6 & 35.3 & .019 & -.073 &  & -225.9  ~ \\
 3$^-$   & 5.62 &     & 140 & .378 & .424 & .39$^{18}$ &   127.4  ~ \\
 5$^-$   & 8.45 &     &  1433 & .071 & .111 & .05$^{18}$ &   133.1  ~ \\
  cpd.   &      &     &     &      &      &      &  -106.5  ~ \\
\end{tabular}
\end{ruledtabular}
\begin{flushleft}
\end{flushleft}
\label{table2}
\end{table}

In Fig. 5, we showed the calculated results for the $5^-$ state that is the highest lying excited state of the 2$^-$ gamma vibrational band and the agreement with the experimental data turned out to be not so good.
This could be explained by the fact that the assignment of the $5^-$ state to the $K^\pi = 2^- $ band is not certain, as the data can be explained as belonging to a $K^\pi = 0^- $ band\cite{18} and it might be necessary to include the couplings with other excited states nearby to describe this state well.
In Table 2, we showed the deformation parameters for the excited states of 2$^-$ gamma vibrational band in $^{20}$Ne and compared with those obtained by nonrelativistic coupled channel calculation. It is shown that the results of Dirac phenomenological calculation agree pretty well with the results of the nonrelativistic calculation for the $3^-$ state, in every case. Also, the changes of the potential strengths and $\chi^{2}$/Ns are given in Table 2 at the six-, eight- and ten-parameter searches. Here, N is the number of experimental cross section data for each state. It is noted that when the 2$^-$ and 3$^-$ states are coupled, the best fit results are obtained. By adding the 3$^-$ state, the $\chi^{2}$/N for the 2$^-$ state is reduced to about 1/3 of the $\chi^{2}$/N that obtained where only the 2$^-$ state is coupled to the ground state, showing that the two step excitation process via channel coupling with the 3$^-$ state is important for the 2$^-$ state excitation. However, the $\chi^{2}$/N for the 3$^-$ state is reduced just a little indicating that the direct step excitation from the ground state is dominant for the 3$^-$ state.

\section{CONCLUSIONS}

Relativistic Dirac coupled channel calculation using optical potential model could describe the high-lying excited states that belong to the 2$^-$ gamma vibrational band at the 800 MeV unpolarized proton inelastic scatterings from an s-d shell nucleus $^{20}$Ne much better than the nonrelativistic coupled channel calculation, especially for the 2$^-$ and 3$^-$ state of the band.
Dirac equations are reduced to the second order differential equations to obtain Schr\"{o}dinger equivalent central and spin-orbit potentials and it is found that surface-peaked phenomena are observed at the real effective central and spin-orbit potentials for the scattering from $^{20}$Ne, as in the case of $^{24}$Mg. The first order vibrational collective models are used to describe the excited states of the 2$^-$ gamma vibrational band in the nucleus and the obtained deformation parameters are compared with those of nonrelativistic calculation. It is found that the deformation parameters of Dirac phenomenological calculation for the 3$^-$ state of the high-lying 2$^-$ gamma vibrational band in $^{20}$Ne agree well with the those of the nonrelativistic calculations using the same Woods-Saxon potential shape.
It is shown that pure direct transition from the ground state is dominant for the 3$^-$ state excitation and the two step excitation via channel coupling with the 3$^-$ state is essential for the $2^-$ state excitation of the 2$^-$ gamma vibrational band in $^{20}$Ne.

\begin{acknowledgments}
This work was supported by the research grant of the Kongju National University in 2013.
\end{acknowledgments}


\begin{references}
\bibitem{1} L. G. Arnold, B. C. Clark, R. L. Mercer, and  P. Swandt, Phys. Rev. C {\bf 23}, 1949 (1981).
\bibitem{2} B. C. Clark, R. L. Mercer, and  P. Swandt, Phys. Lett. {\bf 122B}, 211 (1983).
\bibitem{3} S. Hama, B. C. Clark, R. E. Kozack, S. Shim, E. D. Cooper, R. L. Mercer, and B. D. Serot, Phys. Rev. C {\bf 37}, 1111 (1988).
\bibitem{4} S. Shim, Ph. D. Thesis, The Ohio State University 1989: L. Kurth, B. C. Clark, E. D. Cooper, S. Hama, S. Shim, R. L. Mercer, L. Ray, and G. W. Hoffmann, Phys.  Rev. C {\bf 49}, 2086 (1994).
\bibitem{5} S. Shim, B.C. Clark, E.D. Cooper, S. Hama, R.L. Mercer, L. Ray, J. Raynal, and H.S. Sherif, Phys. Rev. C {\bf 42}, 1592 (1990).
\bibitem{6} R. de Swiniarski, D. L. Pham, and J. Raynal, Z. Phys. A-Hadrons and Nuclei {\bf 343}, 179 (1992).
\bibitem{7} D. L. Pham and R de Swiniarski, Nuovo Cimento A {\bf 107}, (1994) 1405.
\bibitem{8} J. J. Kelly, Phys. Rev. C {\bf71}, 064610 (2005).
\bibitem{9} S. Shim, M. W. Kim, B. C. Clark, and L. Kurth Kerr, Phys. Rev. C {\bf 59}, 317 (1999).
\bibitem{10} S. Shim, Shin-Ho Ryu and Min-Soo Kim, J. Korean. Phys. Soc. {\bf 51}, 271 (2007); S. Shim, Shin-Ho Ryu and Min-Soo Kim, J. Korean. Phys. Soc. {\bf 53}, 1146 (2008).
\bibitem{11} S. Shim and M. W. Kim, Int. Jou. of Mod. Phys. E {\bf 21}, 1250098 (2012).
\bibitem{12} J. Raynal, {\it Computing as a Language of Physics}, ICTP International Seminar Course, 281(IAEA, Italy, 1972): J. Raynal, {\it Notes on ECIS94}, Note CEA-N-2772, 1994.
\bibitem{13} C. J. Horowitz and B. D. Serot, Nucl. Phys. A {\bf 368}, 503 (1981).
\bibitem{14} R. J.  Furnstahl, C. E.  Price, and G.  E. Walker, Phys.  Rev. C  {\bf 36}, 2590 (1987).
\bibitem{15} L. Ray and G. W. Hoffmann, Phys. Rev. C {\bf 31}, 538 (1986).
\bibitem{16} S. Shim and M. W. Kim, J. Korean. Phys. Soc. {\bf 64}, 483 (2014).
\bibitem{17} G. S. Blanpied, G. A. Balchin, G. E. Langston, B. G. Ritchie, M. L. Barlett, G. W. Hoffmann, J. A. McGill, M. A. Franey, M. Gazzaly, B. H. Wildenthal, Phys. Rev. C {\bf 30}, 1233 (1984).
\bibitem{18} G. S. Blanpied, B. G. Ritchie, M. L. Barlett, R. W. Fergerson, G. W. Hoffmann, J. A. McGill, B. H. Wildenthal,  et al Phys. Rev. C {\bf 38}, 2180 (1988).
\end{references}
\end{document}